\documentclass[conference]{IEEEtran}
\IEEEoverridecommandlockouts
\usepackage{cite}
\usepackage{amsmath,amssymb,amsfonts}
\usepackage{algorithmic}
\usepackage{graphicx}
\usepackage{textcomp}
\usepackage{booktabs}
\usepackage{subcaption}
\usepackage[hidelinks]{hyperref}
\usepackage{multirow}
\usepackage{mathtools}
\usepackage{xcolor}
\def\BibTeX{{\rm B\kern-.05em{\sc i\kern-.025em b}\kern-.08em
    T\kern-.1667em\lower.7ex\hbox{E}\kern-.125emX}}

\usepackage{comment}

\begin{document}

\IEEEoverridecommandlockouts
\title{Multi-Node Transformers for Object Detection and Captioning Across Distributed Edge Networks\\
}

\author{\IEEEauthorblockN{Saurabh Kumar}
\IEEEauthorblockA{\textit{Walmart} \\
Sunnyvale, USA \\
saurabh.kumar4@walmart.com}
\and 
\IEEEauthorblockN{Dewant Katare}
\IEEEauthorblockA{\textit{Delft University of Technology} \\
Delft, the Netherlands \\
d.katare@tudelft.nl}
\and 
\IEEEauthorblockN{Prakriti Jain }
\IEEEauthorblockA{\textit{Heidelberg University} \\
Heidelberg, Germany \\
prakriti.jain@stud.uni-heidelberg.de}
}

\maketitle

\begin{abstract}
This paper presents a unified transformer-based framework for object detection
and dense region captioning across distributed edge networks. The architecture
uses a shared visual encoder followed by detection, region extraction, and
autoregressive caption-decoding stages, allowing localization and region
description to reuse the same image features. This design removes the second
image-encoding path required by decoupled detection and captioning pipelines.
To support execution on resource-limited devices, a workload-allocation
formulation assigns model stages and input streams across connected edge nodes
while accounting for execution time, memory capacity, processor utilization,
power limits, and intermediate tensor-transfer cost. The framework is evaluated
on a six-device edge testbed comprising four NVIDIA Jetson Orin Nano and Orin NX
inference nodes and two Raspberry Pi 5 auxiliary CPU nodes, with TensorRT FP16
and INT8 execution on the NVIDIA devices. On a Jetson Orin Nano, the unified
INT8 model achieves $38.1$ FPS with $3.5$ GB peak memory use and $11.8$ W
measured power. Under four concurrent streams, the proposed allocation achieves
$35.2$ FPS and $28.4$ ms mean end-to-end latency with an inter-node transfer
rate of $6.8$ MB/s.
\end{abstract}


\section{Introduction}
\label{sec:introduction}

Edge vision systems increasingly combine spatial perception with semantic
description in applications such as traffic monitoring, industrial inspection,
and autonomous sensing. Object detection provides localized object information,
while region captioning adds textual descriptions for users and downstream
text-based systems. Executing both tasks at the edge can reduce raw-data
transfer and support latency-sensitive operation.

Existing pipelines commonly use separate detection and captioning models,
causing the same image to be encoded more than once. This repeated processing
increases computation, memory use, and energy demand. Lightweight object
detectors have been studied for edge deployment
\cite{al2024fpga,mittal2024comprehensive, katare2023energy}, while recent vision-text systems
address traffic captioning, remote-sensing captioning, and multimodal object
understanding \cite{dinh2024trafficvlm,bazi2024rs,zang2025contextual}. These directions are
usually considered separately rather than within a shared edge-oriented inference graph. Transformer-based vision-text models also remain expensive for low-power hardware because visual encoding, token processing, and autoregressive decoding compete for limited memory and compute resources. Prior work reduces this cost through token sampling, simplified attention, pruning, and quantization \cite{chen2024ta,koohpayegani2024sima,eliopoulos2025pruning,choi2025gradq}.
Distributed inference provides another option by partitioning models across connected devices
\cite{hu2024edge,wen2025easyvit,li2026collaborative, yuan2024edge,chaitra2025scalable,katare2024arasec}. However, the benefit of partitioning depends on where the model is divided: large intermediate tensors increase transfer
time, while detection and autoregressive captioning impose different execution
loads. Static placement therefore perform poorly when stream load or link
conditions change.

This paper presents a unified transformer-based framework for object detection
and dense region captioning across heterogeneous edge nodes. A shared visual
encoder processes each input image once, after which detection, region
extraction, and caption decoding operate on the resulting features. The model
is divided into assignable execution stages, and a communication-aware
allocation formulation selects their placement according to execution time,
memory capacity, processor load, power limits, and inter-node transfer cost.
The framework is deployed with TensorRT FP16 and INT8 and evaluated under
changing stream loads and network conditions on a six-node heterogeneous edge
testbed. The main contributions are:
\begin{itemize}
\item A unified transformer model that shares visual feature extraction
between object detection and dense region captioning.

\item A multi-node allocation formulation for assigning encoding,
detection, region extraction, and caption decoding across heterogeneous
edge nodes while accounting for memory, load, power, and communication
constraints.

\item A TensorRT FP16/INT8 evaluation covering task quality, throughput,
latency, memory use, power consumption, and inter-node traffic under
different stream and network conditions.

\end{itemize}

The paper is organized as follows. Section~\ref{sec:related_work} reviews object detection, dense region captioning, efficient transformer models, and distributed edge inference. Section~\ref{sec:system_model} presents the shared visual architecture, distributed execution model, communication-cost formulation, and workload allocation problem. Section~\ref{sec:experimental_setup} describes the dataset, hardware testbed, deployment settings, baselines, and evaluation metrics. Section~\ref{sec:experimental_results} presents the experimental results and discussion. Section~\ref{sec:conclusion} summarizes the findings and future work.

\section{Related Work}
\label{sec:related_work}

\subsection{Edge Detection and Vision-text Models}

Edge object detection has been studied through lightweight architectures,
hardware-aware implementation, and task offloading
\cite{al2024fpga,mittal2024comprehensive,yuan2024edge}. These approaches reduce
local inference cost but mainly address localization and classification.
Vision-text models extend visual processing with captioning, grounding,
question answering, and object-level reasoning
\cite{dinh2024trafficvlm,bazi2024rs,zang2025contextual,feng2025vision}.
When detection and caption generation are deployed as independent models,
however, each task may use a separate visual encoder, increasing computation
and memory demand. The present work instead shares one visual representation
between localization and dense region captioning.

\subsection{Efficient Transformer Execution}

Vision-transformer execution on edge hardware has been reduced through
attention reformulation, token sampling, pruning, redundancy removal, and
quantization
\cite{koohpayegani2024sima,chen2024ta,eliopoulos2025pruning,
wang2025cst,choi2025gradq}. These methods lower the cost of processing a
single model, but they do not determine how heterogeneous stages such as
visual encoding, detection, region extraction, and autoregressive decoding
should be assigned across several devices. Their execution characteristics
also differ: detection follows a largely fixed computation path, whereas
caption-decoding cost varies with the number of retained regions and generated
tokens.

\subsection{Distributed Transformer Inference}

Distributed inference partitions model execution across connected devices to
reduce the load on individual nodes. Prior studies consider transformer
partitioning, collaborative edge execution, multi-tier inference, and hybrid
data-model parallelism
\cite{hu2024edge,wen2025easyvit,li2026collaborative,chaitra2025scalable,ding2025distributed}. The benefit of model partitioning
depends on both local execution cost and the size of intermediate tensors
transferred between nodes. Large feature tensors can offset the gain from
offloading, particularly under restricted bandwidth.

Existing work does not jointly address shared visual encoding,
variable region-caption decoding, and communication-aware stage placement in
a multi-node edge setting. The framework in this paper combines these elements
by sharing one visual encoder between detection and captioning and selecting
stage placement according to processing time, memory demand, node load, power
limits, and intermediate tensor-transfer cost.

\begin{figure*}[t]
\centering
\includegraphics[width=\textwidth]{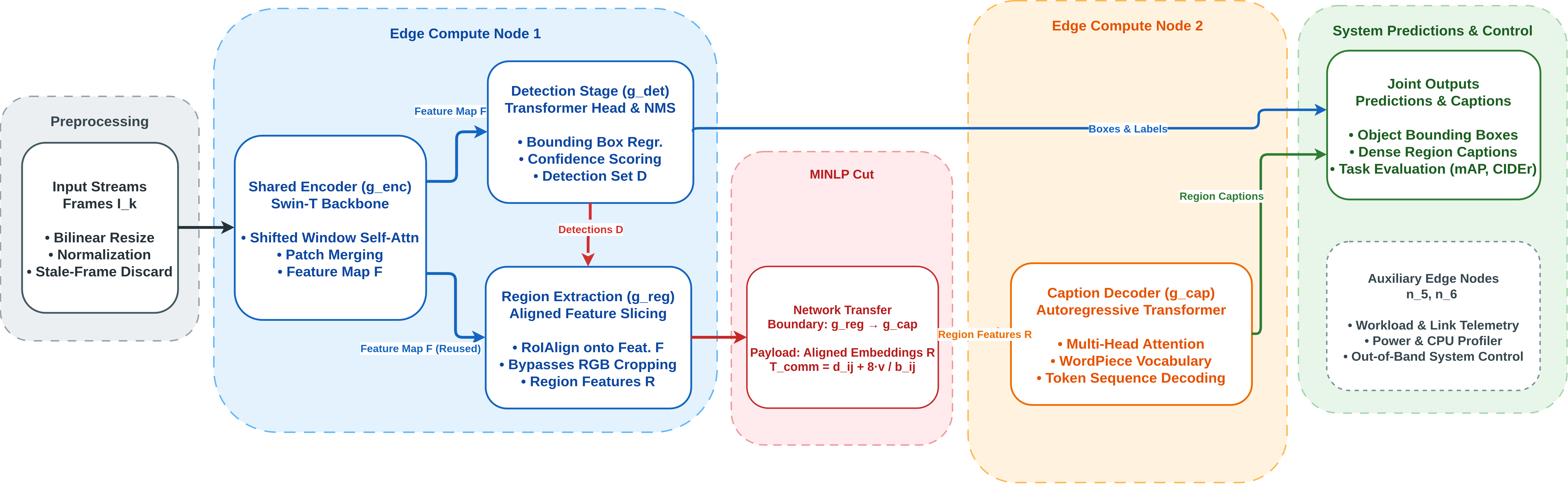}
\caption{End-to-end framework and distributed execution pipeline across heterogeneous edge devices. The shared Swin-T encoder ($g_{\mathrm{enc}}$) produces feature map $\mathbf{F}$, which is reused locally by the detection ($g_{\mathrm{det}}$) and region extraction ($g_{\mathrm{reg}}$) stages on Edge Compute Node 1. The MINLP placement partitions execution at the $g_{\mathrm{reg}} \rightarrow g_{\mathrm{cap}}$ boundary, transferring compact region embeddings $\mathbf{R}$ across the inter-node link to the autoregressive caption decoder ($g_{\mathrm{cap}}$) on Edge Compute Node 2 while auxiliary nodes perform out-of-band telemetry.}
\label{fig:pipeline_architecture}
\end{figure*}

\section{System Model and Problem Formulation}
\label{sec:system_model}

The system processes multiple camera streams across a network of edge devices.
Each frame is encoded once and then passes through detection, region extraction,
and autoregressive region-caption decoding. The allocation problem determines
where the computational stages execute while accounting for processing time,
memory limits, processor load, power limits, and inter-node communication cost.

\subsection{Execution and Placement Model}

Let $\mathcal{N}=\{n_1,\ldots,n_N\}$ denote the set of physical edge devices
and $\mathcal{N}_{\mathrm{cmp}}\subseteq\mathcal{N}$ denote the subset of
devices eligible for computational-stage placement. Let
$\mathcal{S}=\{s_1,\ldots,s_K\}$ denote the set of active streams, where stream
$s_k$ offers frames at rate $f_k$. After stale-frame removal, let
$\tilde{f}_k\leq f_k$ denote the rate admitted to the inference pipeline.
For a compute node $n_i\in\mathcal{N}_{\mathrm{cmp}}$, $m_i$ denotes
available memory, $u_{0,i}$ background processor utilization, and $p_i$ the
configured power limit. The link from node $n_i$ to node $n_j$ has measured
bandwidth $b_{ij}$ and link delay $d_{ij}$.

The inference pipeline contains the stage set $\mathcal{G}=\{g_{\mathrm{enc}},g_{\mathrm{det}},g_{\mathrm{reg}},
g_{\mathrm{cap}}\}$, corresponding to shared encoding, detection, region extraction, and caption decoding, respectively, as illustrated in
Fig.~\ref{fig:pipeline_architecture}. The encoder produces a visual feature tensor $\mathbf{F}$ from input image $\mathbf{I}$. Detection produces retained detections $\mathcal{D}$ containing region locations and associated predictions. Region extraction obtains aligned region features $\mathbf{R}$ directly from $\mathbf{F}$ and $\mathcal{D}$ on the local device, and the autoregressive decoder generates one caption for each retained region. Because region features are sliced directly from $\mathbf{F}$, the captioning branch avoids a second image crop and redundant visual encoding stage. 

The captioning branch uses a two-layer autoregressive Transformer decoder with
hidden dimension $d=256$ and four attention heads. Region features
$\mathbf{R}$ condition the decoder, which predicts tokens over a WordPiece
vocabulary of $V=30{,}522$. The decoder is trained using token-level
cross-entropy loss.

Binary variable $x_{g,i}\in\{0,1\}$ indicates whether stage $g$ is assigned
to compute node $n_i\in\mathcal{N}_{\mathrm{cmp}}$, while
$z_{k,i}\in\{0,1\}$ specifies the ingress node for stream $s_k$. A common
computational-stage placement is used by all active streams, while
$z_{k,i}$ specifies the ingress assignment of each stream. Each computational
stage receives exactly one assignment. If $q_g$ denotes the memory required by
the parameters and execution buffers of stage $g$, a placement is feasible only
when the cumulative memory of stages assigned to node $n_i$ does not exceed
$m_i$.

\subsection{Caption-Aware Processing Cost}

The encoder, detector, and region-extraction stages have profiled mean
per-frame execution times $\bar{\tau}_{g,i}$ on compute node $n_i$.
Caption-decoding runtime varies with both the number of retained regions and
the generated sequence length. For sequential region decoding, the mean
caption-stage service time on node $n_i$ is modeled as
\begin{equation}
\bar{\tau}_{\mathrm{cap},i}
=
\tau_{\mathrm{init},i}(M_{\mathrm{avg}})
+
M_{\mathrm{avg}}T_{\mathrm{avg}}\tau_{\mathrm{tok},i},
\label{eq:caption_time_estimator}
\end{equation}
where $M_{\mathrm{avg}}$ is the profiled mean retained-region count,
$T_{\mathrm{avg}}$ is the mean caption length in tokens,
$\tau_{\mathrm{init},i}$ is the region-conditioning latency, and
$\tau_{\mathrm{tok},i}$ is the mean token-decoding latency.

The aggregate service time assigned to node $n_i$ is
\begin{equation}
\bar{\tau}_i
=
\sum_{g\in\mathcal{G}}x_{g,i}\bar{\tau}_{g,i}.
\end{equation}

Since only admitted frames enter the computational pipeline, node utilization
is estimated using the aggregate admitted processing rate:
\begin{equation}
\rho_i
=
u_{0,i}
+
\left(
\sum_{k=1}^{K}\tilde{f}_k
\right)
\bar{\tau}_i,
\qquad
n_i\in\mathcal{N}_{\mathrm{cmp}}.
\label{eq:node_utilization}
\end{equation}

The distinction between $f_k$ and $\tilde{f}_k$ separates offered workload
from processed workload when the input rate exceeds available processing
capacity. This is consistent with the stale-frame removal policy used in the
experiments.

To account for increasing delay as utilization approaches capacity, the node
residence-time cost is approximated as
\begin{equation}
T_{\mathrm{node},i}
=
\frac{\bar{\tau}_i}{1-\rho_i},
\qquad
\rho_i\leq\rho_{\max}<1.
\label{eq:node_residence_time}
\end{equation}
This expression is used as a placement-cost approximation rather than an exact
queueing result and is validated against measured node delays in
Section~\ref{sec:experimental_results}.

\subsection{Communication and End-to-End Cost}

The payload transmitted across consecutive stage boundaries depends on the
selected placement. An ingress transfer carries the input frame, a
$g_{\mathrm{enc}}\rightarrow g_{\mathrm{det}}$ boundary transfers the shared
feature tensor $\mathbf{F}$, a
$g_{\mathrm{det}}\rightarrow g_{\mathrm{reg}}$ boundary transfers serialized
detection records $\mathcal{D}$, and a
$g_{\mathrm{reg}}\rightarrow g_{\mathrm{cap}}$ boundary transfers aligned
region features $\mathbf{R}$.

If $v_{g,h}$ denotes the payload size between consecutive stages $g$ and $h$,
the inter-node transfer time from node $n_i$ to node $n_j$ is
\begin{equation}
T_{ij}^{g,h}
=
d_{ij}
+
\frac{8v_{g,h}}{b_{ij}}.
\label{eq:stage_transfer_latency}
\end{equation}
The factor $8$ converts payload size from bytes to bits. As shown in
Fig.~\ref{fig:pipeline_architecture}, partitioning across heterogeneous devices
(e.g., placing $\{g_{\mathrm{enc}}, g_{\mathrm{det}}, g_{\mathrm{reg}}\}$ on
Node~$1$ and $g_{\mathrm{cap}}$ on Node~$2$) restricts the transferred payload
across the inter-node link to the compact region embeddings $\mathbf{R}$,
avoiding the prohibitive bandwidth cost of transmitting raw feature maps
$\mathbf{F}$.

The total inter-stage communication cost $T_{\mathrm{comm}}$ is the sum of
transfer times across consecutive stage boundaries assigned to different
compute nodes. Similarly, $R_{\mathrm{trans}}$ denotes the aggregate rate of
input and intermediate payloads transmitted between nodes for the admitted
stream workload.

For stream $s_k$, end-to-end processing latency combines ingress transfer,
node residence times, inter-stage communication, and pipeline synchronization:
\begin{equation}
T_{\mathrm{e2e},k}
=
T_{\mathrm{in},k}
+
\sum_{i:n_i\in\mathcal{N}_{\mathrm{cmp}}}
T_{\mathrm{node},i}
+
T_{\mathrm{comm}}
+
T_{\mathrm{sync}}.
\label{eq:end_to_end_time}
\end{equation}

The optimization considers the mean end-to-end latency
\begin{equation}
\bar{T}_{\mathrm{e2e}}
=
\frac{1}{K}
\sum_{k=1}^{K}T_{\mathrm{e2e},k}.
\end{equation}

Node power is estimated from idle and peak measurements as
\begin{equation}
P_i(\mathbf{X})
=
P_{\mathrm{idle},i}
+
\rho_i
\left(
P_{\mathrm{peak},i}-P_{\mathrm{idle},i}
\right).
\end{equation}
This linear relation is used only as a first-order surrogate within the
placement objective; experimentally reported power values are obtained from
direct device measurements.

\subsection{Workload Allocation}

To limit uneven resource use, the allocation penalizes memory and processor
load differences across participating compute nodes. Let
$\mathcal{N}_{\mathrm{act}}\subseteq\mathcal{N}_{\mathrm{cmp}}$ denote the
compute nodes assigned at least one computational stage. Processor-load
imbalance is quantified by
\begin{equation}
U_{\mathrm{load}}
=
\max_{i:n_i\in\mathcal{N}_{\mathrm{act}}}\rho_i
-
\min_{i:n_i\in\mathcal{N}_{\mathrm{act}}}\rho_i,
\label{eq:load_imbalance}
\end{equation}
with $U_{\mathrm{mem}}$ defined analogously using the occupied-memory fraction
of each active compute node.

Let $\mathbf{X}=[x_{g,i}]$ and $\mathbf{Z}=[z_{k,i}]$ collect the stage and
stream assignments, respectively. The objective jointly minimizes normalized
mean latency, network-transfer rate, memory imbalance, and processor-load
imbalance:
\begin{equation}
\min_{\mathbf{X},\mathbf{Z}}
\quad
\mathcal{J}
=
\alpha
\frac{\bar{T}_{\mathrm{e2e}}}{T_{\mathrm{ref}}}
+
\beta
\frac{R_{\mathrm{trans}}}{R_{\mathrm{ref}}}
+
\gamma U_{\mathrm{mem}}
+
\delta U_{\mathrm{load}}.
\label{eq:allocation_objective}
\end{equation}

The placement is subject to
\begin{equation}
\begin{aligned}
\sum_{i:n_i\in\mathcal{N}_{\mathrm{cmp}}}x_{g,i}
&=1,
&&\forall g\in\mathcal{G},
\\
\sum_{i=1}^{N}z_{k,i}
&=1,
&&\forall s_k\in\mathcal{S},
\\
\sum_{g\in\mathcal{G}}x_{g,i}q_g
&\leq m_i,
&&\forall n_i\in\mathcal{N}_{\mathrm{cmp}},
\\
\rho_i
&\leq\rho_{\max},
&&\forall n_i\in\mathcal{N}_{\mathrm{cmp}},
\\
P_i(\mathbf{X})
&\leq p_i,
&&\forall n_i\in\mathcal{N}_{\mathrm{cmp}}.
\end{aligned}
\label{eq:allocation_constraints}
\end{equation}

The variables satisfy
$x_{g,i},z_{k,i}\in\{0,1\}$. The normalization constants are fixed to
$T_{\mathrm{ref}}=100$ ms and $R_{\mathrm{ref}}=50$ MB/s for all
experiments. These constants scale the latency and network-transfer terms
before application of the objective weights.

The binary assignments, utilization-dependent residence time, and
placement-dependent communication cost result in a mixed-integer nonlinear
program (MINLP). The allocation problem is implemented in Pyomo and solved
using SCIP 8.0. The objective weights are fixed to
$\alpha=0.40$, $\beta=0.30$, $\gamma=0.15$, and $\delta=0.15$ for all
experiments. SCIP is configured with a $5.0$~s time limit and a relative
optimality-gap target of $1\%$. The best feasible incumbent is used when the
time limit is reached before the target gap.

\section{Experimental Setup and Evaluation}
\label{sec:experimental_setup}

The evaluation examines task-level prediction quality and system-level
execution behavior under different precision modes, stream loads, network
conditions, and stage placements. Experiments use the Visual Genome benchmark
and a six-device edge testbed comprising four NVIDIA Jetson inference nodes
and two Raspberry Pi 5 auxiliary CPU nodes.

\subsection{Dataset and Workload Profiling}

The experiments use the Visual Genome dataset
\cite{krishna2017visual}, which provides localized object annotations and
natural-text region descriptions for joint localization and dense
region-caption evaluation. We use an image-level $80/10/10$ split for
training, validation, and testing. All annotations associated with an image
remain within the same subset to prevent image-level overlap.

Input images are resized to $640\times640$ using bilinear interpolation and
normalized with ImageNet channel statistics. Color jitter and random
horizontal flipping are applied during training. Region descriptions are
processed using WordPiece tokenization with a maximum sequence length of
$T_{\max}=16$ tokens; longer descriptions are truncated and shorter sequences
are padded.

The allocation model in Section~\ref{sec:system_model} uses profiling
measurements obtained from the validation subset and target devices. After
applying a detection confidence threshold of $0.5$ and non-maximum suppression
with an IoU threshold of $0.6$, the mean retained-region count is
$M_{\mathrm{avg}}=14.2$, while the mean caption length is
$T_{\mathrm{avg}}=8.5$ tokens. For each profiled stage--device pair, mean
execution time $\bar{\tau}_{g,i}$ is measured over $500$ timed executions
following $50$ warm-up iterations. TensorRT FP16 and INT8 profiling is
performed on the NVIDIA devices, while Raspberry Pi 5 measurements use
CPU execution.

Concurrent workloads are generated by replaying validation images as
$K\in\{2,4,6,8\}$ independent streams. Each stream operates at a configured
rate selected from $f_k\in\{10,15,20,30\}$ FPS. These replay streams are used
to vary aggregate system load; no temporal modeling or frame-to-frame
consistency is assumed.

\subsection{Testbed and Deployment Configuration}

The physical testbed contains six edge devices, summarized in
Table~\ref{tab:testbed_configuration}. Jetson Orin Nano and Orin NX devices
provide GPU-accelerated inference, while the Raspberry Pi 5 devices serve as
auxiliary CPU edge nodes for runtime profiling and system monitoring. Their
stage measurements are also used to characterize when CPU-only execution
would violate the utilization limits of the placement model.

\begin{table}[t]
\caption{Six-device edge testbed configuration.}
\label{tab:testbed_configuration}
\centering
\small
\setlength{\tabcolsep}{3pt}
\begin{tabular}{lccc}
\toprule
\textbf{Nodes} & \textbf{Device} & \textbf{Memory} & \textbf{Role/Mode} \\
\midrule
$n_1,n_2$ & Jetson Orin Nano & 8 GB & INT8/FP16, 15 W \\
$n_3,n_4$ & Jetson Orin NX & 16 GB & INT8/FP16, 25 W \\
$n_5,n_6$ & Raspberry Pi 5 & 8 GB & CPU auxiliary \\
\bottomrule
\end{tabular}
\end{table}

The NVIDIA devices run JetPack 6.2 with Jetson Linux 36.4.3 and TensorRT
10.3. Raspberry Pi 5 devices run the 64-bit Raspberry Pi operating system and
use ONNX Runtime 1.18 on the ARM64 CPU with NEON acceleration. The placement
policy evaluates measured execution time, available memory, background
utilization, configured power limits, and link conditions rather than assigning
stages according to fixed hardware tiers.

Table~\ref{tab:stage_profile} reports representative execution times for the
four computational stages. The Raspberry Pi measurements characterize the
cost of CPU-only execution relative to the accelerated Jetson platforms.

\begin{table}[t]
\caption{Representative per-stage execution times ($\bar{\tau}_{g,i}$) across edge devices.}
\label{tab:stage_profile}
\centering
\small
\setlength{\tabcolsep}{3pt}
\begin{tabular}{lcccc}
\toprule
\textbf{Device} &
\textbf{$g_{\mathrm{enc}}$} &
\textbf{$g_{\mathrm{det}}$} &
\textbf{$g_{\mathrm{reg}}$} &
\textbf{$g_{\mathrm{cap}}$} \\
\midrule
Jetson Orin NX (INT8) &
$5.4$ ms & $2.3$ ms & $1.0$ ms & $3.8$ ms \\

Jetson Orin Nano (INT8) &
$11.2$ ms & $4.8$ ms & $2.1$ ms & $8.1$ ms \\

Raspberry Pi 5 (CPU) &
$148.5$ ms & $38.5$ ms & $14.2$ ms & $42.0$ ms \\
\bottomrule
\end{tabular}
\end{table}

The Raspberry Pi 5 can execute the profiled stages on the CPU, but its service
times are substantially higher than those of the INT8 Jetson devices.
Region extraction is the least costly profiled Pi stage at $14.2$ ms,
compared with $38.5$ ms for detection, $42.0$ ms for caption decoding, and
$148.5$ ms for shared encoding. For a given workload, a stage--device
combination is considered infeasible when its predicted utilization violates
the constraint $\rho_i\leq\rho_{\max}$. The Pi measurements therefore provide
a CPU-edge reference and allow infeasible CPU placements to be screened as
the aggregate stream rate increases. The high-throughput placement experiments
reported in Section~\ref{sec:experimental_results} use the GPU-accelerated
NVIDIA nodes for computational-stage execution.

Stages assigned to NVIDIA devices are exported to ONNX using opset 17 and
compiled with TensorRT 10.3. Both FP16 and INT8 execution are evaluated.
INT8 engines use explicit \texttt{QuantizeLinear} and
\texttt{DequantizeLinear} nodes inserted during post-training quantization.
Quantization parameters are estimated from $1000$ unlabeled images selected
from the training subset, with no overlap with validation or test data.
Operations without an INT8 implementation execute using a supported
higher-precision kernel, and the resulting layer precisions are recorded
during engine construction.

The nodes communicate through a managed $1~\mathrm{Gbps}$ Ethernet switch.
Controlled link conditions are configured using \texttt{tc-netem}. Evaluated
bandwidths are
$b_{ij}\in\{10,50,100,500,1000\}~\mathrm{Mbps}$, with added delays
$d_{ij}\in\{1,5,15,30,50\}~\mathrm{ms}$. Each experimental configuration is
repeated $10$ times under identical stream, precision, and network settings.

\subsection{Baselines and Evaluation Metrics}

The proposed allocation policy is evaluated against four placement baselines.
\textit{Single-node execution} places all computational stages on one feasible
NVIDIA node. \textit{Static pipeline placement} uses a fixed assignment across
the available NVIDIA compute nodes and does not change with stream or network
conditions. \textit{Greedy load-based placement} assigns each stage to the
feasible compute node with the lowest measured utilization, while
\textit{memory-guided placement} uses available memory without accounting for
intermediate tensor size or link bandwidth. All placement methods use the same
model parameters, precision mode, input streams, detection thresholds, NMS
settings, and caption-decoding procedure.

At the model level, the proposed shared-encoder architecture is compared with
a decoupled baseline containing a Swin-B detector and an independent ViT-B
caption encoder. Both systems use the same training and evaluation subsets.
Because the architectures have different encoder capacities, this comparison
measures end-to-end deployment cost rather than attributing the full difference
to encoder sharing alone.

Localization quality is reported using $\mathrm{mAP}_{50}$, while
region-caption quality is measured using CIDEr. System measurements include
end-to-end latency, completed-frame throughput, inter-node transfer rate,
peak per-node memory use, node utilization, and power consumption.
End-to-end latency is measured from frame availability at the ingress node
until detection and all retained region captions are complete. Throughput
counts frames for which the complete pipeline finishes per second. Network
traffic includes both ingress and intermediate transfers between compute
nodes.

Jetson memory measurements refer to shared system memory, and node utilization
refers to measured GPU utilization on the NVIDIA compute nodes. Raspberry Pi
CPU utilization and process memory are used only for auxiliary profiling and
feasibility characterization. Modeled power used by the allocation objective
is kept separate from direct device measurements. The terms
$U_{\mathrm{mem}}$ and $U_{\mathrm{load}}$ represent max--min differences
across active computational nodes and are not interpreted as statistical
variances.

The evaluation addresses three questions: whether shared encoding and
reduced-precision execution lower deployment cost while preserving task
quality; how bandwidth, link delay, and concurrent stream load affect
distributed execution; and whether the proposed allocation improves latency,
network traffic, and load distribution relative to the placement baselines.

\section{Experimental Results and Discussion}
\label{sec:experimental_results}

The evaluation examines model efficiency, placement-policy performance,
sensitivity to network bandwidth, stream-load scaling, and the accuracy of the
utilization-dependent residence-time model. The Raspberry Pi 5 nodes are used
as lightweight control and monitoring endpoints in the testbed; the
high-throughput stage-placement results reported in this section use the
GPU-accelerated NVIDIA nodes.

\subsection{Model Efficiency}

Table~\ref{tab:shared_encoder_results} compares the proposed unified model
with the decoupled dual-backbone baseline on a single Jetson Orin Nano node.
The proposed model uses a shared Swin-T encoder and contains $48.2$ million
parameters, whereas the decoupled baseline uses a Swin-B detector with
$88$ million parameters and an independent ViT-B caption encoder with
$86$ million parameters. Because the systems differ in both feature-sharing
structure and encoder capacity, the comparison represents their complete
deployment cost rather than the isolated effect of encoder sharing.

\begin{table}[t]
\caption{Single-node performance across model and precision configurations
($K=1$, node $n_1$).}
\label{tab:shared_encoder_results}
\centering
\small
\setlength{\tabcolsep}{3.1pt}
\begin{tabular}{lcccccc}
\toprule
\textbf{Architecture} &
\textbf{Mode} &
$\mathbf{mAP}_{50}$ &
\textbf{CIDEr} &
\textbf{FPS} &
\textbf{Mem.} &
\textbf{Power} \\
\midrule
Decoupled baseline &
FP16 &
31.4 &
94.2 &
14.2 &
6.2 GB &
14.8 W \\

Proposed unified &
FP16 &
31.1 &
93.8 &
22.5 &
3.9 GB &
13.2 W \\

Proposed unified &
INT8 &
30.6 &
92.1 &
38.1 &
3.5 GB &
11.8 W \\
\bottomrule
\end{tabular}
\end{table}

The unified FP16 model increases throughput from $14.2$ to $22.5$ FPS while
reducing peak memory from $6.2$ to $3.9$ GB. The corresponding differences in
$\mathrm{mAP}_{50}$ and CIDEr are $0.3$ and $0.4$ points, respectively.
INT8 execution further increases throughput from $22.5$ to $38.1$ FPS,
corresponding to a $1.69\times$ increase over unified FP16 execution. Peak
memory decreases from $3.9$ to $3.5$ GB and measured power draw decreases from
$13.2$ to $11.8$ W. Relative to unified FP16, INT8 reduces
$\mathrm{mAP}_{50}$ by $0.5$ percentage points and CIDEr by $1.7$ points.
INT8 therefore increases throughput by $69.3\%$ over unified FP16 while
incurring a limited reduction in localization and caption quality.

\subsection{Placement-Policy Performance}

Table~\ref{tab:placement_results} compares the placement methods using four
concurrent streams, each configured at $30$ FPS, giving an offered load of
$120$ FPS. Network bandwidth is fixed at $100$ Mbps with an added link delay
of $5$ ms. Throughput counts frames for which detection and all retained
region captions are completed. When the offered rate exceeds processing
capacity, stale replay frames are discarded and the most recent available
frame is processed; the same policy is applied to all placement methods.

\begin{table}[t]
\caption{Placement performance under four concurrent streams
($b_{ij}=100$ Mbps, $d_{ij}=5$ ms, INT8).}
\label{tab:placement_results}
\centering
\small
\setlength{\tabcolsep}{2.5pt}
\begin{tabular}{lccccc}
\toprule
\textbf{Strategy} &
\textbf{Latency} &
\textbf{FPS} &
\textbf{Traffic} &
\textbf{Memory} &
\textbf{Power} \\
\midrule
Single node &
88.4 ms &
11.3 &
0.0 MB/s &
7.8 GB &
14.9 W \\

Static pipeline &
42.1 ms &
23.8 &
14.2 MB/s &
4.1 GB &
13.5 W \\

Greedy load-based &
36.5 ms &
27.4 &
18.6 MB/s &
4.0 GB &
13.0 W \\

Memory-guided &
49.2 ms &
20.3 &
28.1 MB/s &
3.6 GB &
12.8 W \\

\textbf{Proposed MINLP} &
\textbf{28.4 ms} &
\textbf{35.2} &
\textbf{6.8 MB/s} &
\textbf{3.7 GB} &
\textbf{12.1 W} \\
\bottomrule
\end{tabular}
\end{table}

Single-node execution reaches $7.8$ GB peak memory, leaving limited capacity
for other system and runtime activity. Under the four-stream workload, it
achieves $11.3$ FPS with a mean latency of $88.4$ ms. The proposed allocation
reduces mean latency to $28.4$ ms and achieves $35.2$ FPS, with measured
latencies ranging from $26$ to $31$ ms across repeated runs. Throughput is
$47.9\%$ higher than greedy load-based placement and $73.4\%$ higher than
memory-guided placement.

For this configuration, the selected assignment partitions the execution path
at the $g_{\mathrm{det}}\rightarrow g_{\mathrm{reg}}$ boundary between
GPU-accelerated edge nodes. Serialized detection records at this boundary are
smaller than the encoder feature tensor and aligned region features, reducing
communication demand. The resulting transfer rate is $6.8$ MB/s, compared
with $18.6$ MB/s for greedy placement and $28.1$ MB/s for memory-guided
placement. The power values in Table~\ref{tab:placement_results} correspond to
the measured draw of the Jetson node carrying the largest computational load.

The Raspberry Pi 5 nodes are not assigned compute-intensive stages in this
high-rate configuration. Their measured CPU execution times are used during
profiling to identify stage--device combinations that violate the configured
utilization constraint before placement optimization.

\subsection{Effect of Network Bandwidth}

Table~\ref{tab:bandwidth_sensitivity} reports mean end-to-end latency as
inter-node bandwidth is varied. The stream count, INT8 precision, and added
link delay remain fixed.

\begin{table}[t]
\caption{Effect of inter-node bandwidth on mean latency
($K=4$, $d_{ij}=5$ ms, INT8).}
\label{tab:bandwidth_sensitivity}
\centering
\small
\setlength{\tabcolsep}{3.5pt}
\begin{tabular}{lcccc}
\toprule
\textbf{Bandwidth} &
\textbf{Single} &
\textbf{Greedy} &
\textbf{Memory} &
\textbf{Proposed} \\
\midrule
10 Mbps &
88.4 ms &
142.1 ms &
188.5 ms &
\textbf{52.3 ms} \\

50 Mbps &
88.4 ms &
51.2 ms &
64.0 ms &
\textbf{31.1 ms} \\

100 Mbps &
88.4 ms &
36.5 ms &
49.2 ms &
\textbf{28.4 ms} \\

1000 Mbps &
88.4 ms &
26.1 ms &
27.8 ms &
\textbf{24.2 ms} \\
\bottomrule
\end{tabular}
\end{table}

Single-node latency remains unchanged because no intermediate tensors are
transferred between devices. Distributed placements are more sensitive to
bandwidth because part of their end-to-end latency depends on inter-node
payload transfer.

At $10$ Mbps, the proposed policy reduces mean latency by $63.2\%$ relative
to greedy placement and by $72.3\%$ relative to memory-guided placement.
As bandwidth increases, the difference between the distributed strategies
decreases because communication contributes a smaller fraction of total
latency. The proposed policy maintains the lowest latency among the compared
strategies at each reported bandwidth, showing the importance of intermediate
payload size when link capacity is restricted.

\subsection{Stream Load and Residence-Time Model}

The residence-time approximation in
Eq.~\eqref{eq:node_residence_time} is evaluated by comparing predicted and
measured node delays over utilization levels from $0.10$ to $0.88$. Across
feasible stage--node combinations, the model achieves a mean absolute error of
$2.34$ ms, supporting its use as a placement-cost approximation within the
evaluated utilization range. The experiments use
$\rho_{\max}=0.90$, which prevents placements from approaching the singular
point of Eq.~\eqref{eq:node_residence_time} and leaves capacity for runtime
variation.

The frame-drop rate is defined as
$\text{Drop (\%)} = 100 \times \left(1 -
\frac{f_{\mathrm{completed}}}{f_{\mathrm{offered}}}\right)$,
representing the fraction of offered frames that are not completed by the
pipeline. Table~\ref{tab:stream_scaling} reports system performance as the
number of concurrent streams increases from $K=2$ to $K=8$, with each stream
configured at $30$ FPS ($b_{ij}=100$ Mbps, $d_{ij}=5$ ms, INT8).

\begin{table}[t]
\caption{System performance under increasing concurrent stream load.}
\label{tab:stream_scaling}
\centering
\small
\setlength{\tabcolsep}{5pt}
\begin{tabular}{cccccc}
\toprule
$K$ & \textbf{Offered} & \textbf{Completed} & \textbf{Latency} &
\textbf{Drop} & \textbf{Peak $\rho$} \\
    & \textbf{(FPS)}   & \textbf{(FPS)}     & \textbf{(ms)}    &
\textbf{(\%)} & \\
\midrule
2 & 60  & 58.4 & 21.2 & 2.7\%  & 0.42 \\
4 & 120 & 35.2 & 28.4 & 70.7\% & 0.68 \\
6 & 180 & 38.6 & 34.1 & 78.6\% & 0.81 \\
8 & 240 & 39.1 & 41.5 & 83.7\% & 0.89 \\
\bottomrule
\end{tabular}
\end{table}

At $K=2$, the system completes most of the offered workload, reaching
$58.4$ FPS with a $2.7\%$ drop rate. As the offered load increases, completed
throughput approaches a saturation region of approximately $35$--$39$ FPS,
while mean latency rises from $28.4$ ms at $K=4$ to $41.5$ ms at $K=8$.
Peak utilization increases from $0.42$ to $0.89$ and remains below the
configured limit of $\rho_{\max}=0.90$. The increasing drop rate reflects
offered load beyond the sustained processing capacity rather than a violation
of the configured utilization constraint.

Overall, the unified INT8 model reaches $38.1$ FPS on a single Orin Nano,
while the proposed distributed placement achieves $35.2$ FPS and $28.4$ ms
mean latency under four concurrent streams. At $10$ Mbps, communication-aware
placement reduces mean latency by $63.2\%$ relative to greedy placement and by
$72.3\%$ relative to memory-guided placement.

\section{Conclusion and Future Work}
\label{sec:conclusion}

This paper discussed a unified transformer-based framework for object
detection and dense region captioning across distributed edge nodes. A shared
visual encoder eliminates repeated image encoding, while a workload-allocation
formulation assigns encoding, detection, region extraction, and caption
decoding according to execution time, memory capacity, processor utilization,
power limits, and inter-node transfer cost.

The experimental platform contained four GPU-accelerated NVIDIA edge nodes and
two Raspberry Pi 5 auxiliary CPU nodes used for profiling and feasibility
characterization. TensorRT FP16 and INT8 execution was evaluated on the
Jetson Orin Nano and Orin NX devices. On a Jetson Orin Nano, the unified INT8
model achieved $38.1$ FPS with $3.5$ GB peak memory use and $11.8$ W measured
power. Under four concurrent streams, the proposed placement achieved
$35.2$ FPS and $28.4$ ms mean end-to-end latency with an inter-node transfer
rate of $6.8$ MB/s. At $10$ Mbps bandwidth, mean latency was reduced by
$63.2\%$ relative to greedy placement and by $72.3\%$ relative to
memory-guided placement.

These results support the use of shared visual encoding, reduced-precision
execution, and communication-aware stage placement for joint object detection
and dense region captioning on resource-constrained edge hardware. Future work
will consider online stage migration, node-failure recovery,
feature-transfer compression, and larger heterogeneous edge--cloud
deployments.
\bibliographystyle{IEEEtran}
\bibliography{ref.bib}

\vspace{12pt}

\end{document}